\documentclass[conference]{IEEEtran}
\IEEEoverridecommandlockouts

\usepackage{cite}
\usepackage{amsmath,amssymb,amsfonts}
\usepackage{graphicx}
\usepackage[ruled,vlined]{algorithm2e}
\usepackage{textcomp}
\usepackage{xcolor}
\usepackage{multicol}
\usepackage{url}
\usepackage{mdframed}
\usepackage{listings}
\usepackage{xcolor}
\usepackage{algpseudocode}

\usepackage{mathtools}
\usepackage[hidelinks]{hyperref}
\usepackage{cleveref}

\begin{document}
\bstctlcite{IEEEexample:BSTcontrol}
\title{FAIR Digital Objects for the Realization of Globally Aligned Data Spaces \\
\thanks{© 2024 IEEE. Personal use of this material is permitted. Permission from IEEE must be obtained for all other uses, in any current or future media, including reprinting/republishing this material for advertising or promotional purposes, creating new collective works, for resale or redistribution to servers or lists, or reuse of any copyrighted component of this work in other works. This project is funded by the Helmholtz Metadata Collaboration Platform (HMC), NFDI4Ing (DFG – project number 442146713), and supported by the research program “Engineering Digital Futures” of the Helmholtz Association of German Research Centers.}
}

\author{
        \IEEEauthorblockN{1\textsuperscript{st} Nicolas Blumenröhr \\ \textit{Karlsruhe Institute} \\ \textit{of Technology} \\
 Karlsruhe, Germany \\ 0009-0007-0235-4995 }
        \and
        \IEEEauthorblockN{2\textsuperscript{nd} Philipp-Joachim Ost \\ \textit{Karlsruhe Institute} \\ \textit{of Technology} \\
 Karlsruhe, Germany \\ 0000-0002-7198-0566}
        \and
        \IEEEauthorblockN{3\textsuperscript{rd} Felix Kraus \\ \textit{Karlsruhe Institute} \\ \textit{of Technology} \\
 Karlsruhe, Germany \\ 0000-0002-2102-4170}
        \and
        \IEEEauthorblockN{4\textsuperscript{th} Achim Streit \\ \textit{Karlsruhe Institute} \\ \textit{of Technology} \\
 Karlsruhe, Germany \\ 0000-0002-5065-469X}
    }

\maketitle

\begin{abstract}
The FAIR principles are globally accepted guidelines for improved data management practices with the potential to align data spaces on a global scale. In practice, this is only marginally achieved through the different ways in which organizations interpret and implement these principles.
The concept of FAIR Digital Objects provides a way to realize a domain-independent abstraction layer that could solve this problem, but its specifications are currently diverse, contradictory, and restricted to semantic models. In this work, we introduce a rigorously formalized data model with a set of assertions using formal expressions to provide a common baseline for the implementation of FAIR Digital Objects. The model defines how these objects enable machine-actionable decisions based on the principles of abstraction, encapsulation, and entity relationship to fulfill FAIR criteria for the digital resources they represent. We provide implementation examples in the context of two use cases and explain how our model can facilitate the (re)use of data across domains. We also compare how our model assertions are met by FAIR Digital Objects as they have been described in other projects. Finally, we discuss our results' adoption criteria, limitations, and perspectives in the big data context. Overall, our work represents an important milestone for various communities working towards globally aligned data spaces through FAIRification.
\end{abstract}

\begin{IEEEkeywords}
FAIR Digital Objects, Metadata, Data Spaces, Big Data
\end{IEEEkeywords}

\section{Introduction and Problem Description}
A major portion of scientific work is often related to state-of-the-art analysis; not only in aspects of literature, but also data investigation\cite{wilkinson_fair_2016, wittenburg_digital_2019}. While the application of data resources depends on the specific use case, prior retrieval of administrative information and content assessment is a general prerequisite for data selection. For example, the evaluation of licence information or the preview of image thumbnails are typical tasks for the pre-selection of useful data. Ideally, such tasks are at least in part performed automatically to reduce user workload. There exists, however, a wide variety of data formats, access protocols, storage systems, and associated technologies such as metadata standards, vocabularies, and tooling\cite{wilkinson_interoperability_2017}. The term "data space" was defined in many works as summarized in \cite{curry_data_2022}; based on these definitions and other descriptions, such as from the International Data Spaces Association\footnote{\url{https://internationaldataspaces.org}}, we define a data space as an enclosed environment that contains digital resources (data) that are shared and used by organizations. However, the acquisition, inspection, and analysis of data resources \textit{across} data spaces is a laborious task, mainly due to interoperability issues that typically stem from a combination of technical, semantic, and governance challenges \cite{wilkinson_interoperability_2017}. In the context of big data, these interoperability issues are amplified by the properties of the '5 Vs' - volume, variety, velocity, veracity, and value\cite{delgado_interoperability_2021}. This has a negative impact on the progress in data-intensive fields, for example in Artificial Intelligence (AI) research and application\cite{huerta_fair_2023}.

\begin{figure}[tbp]
    \centering
    \includegraphics[width=\columnwidth]{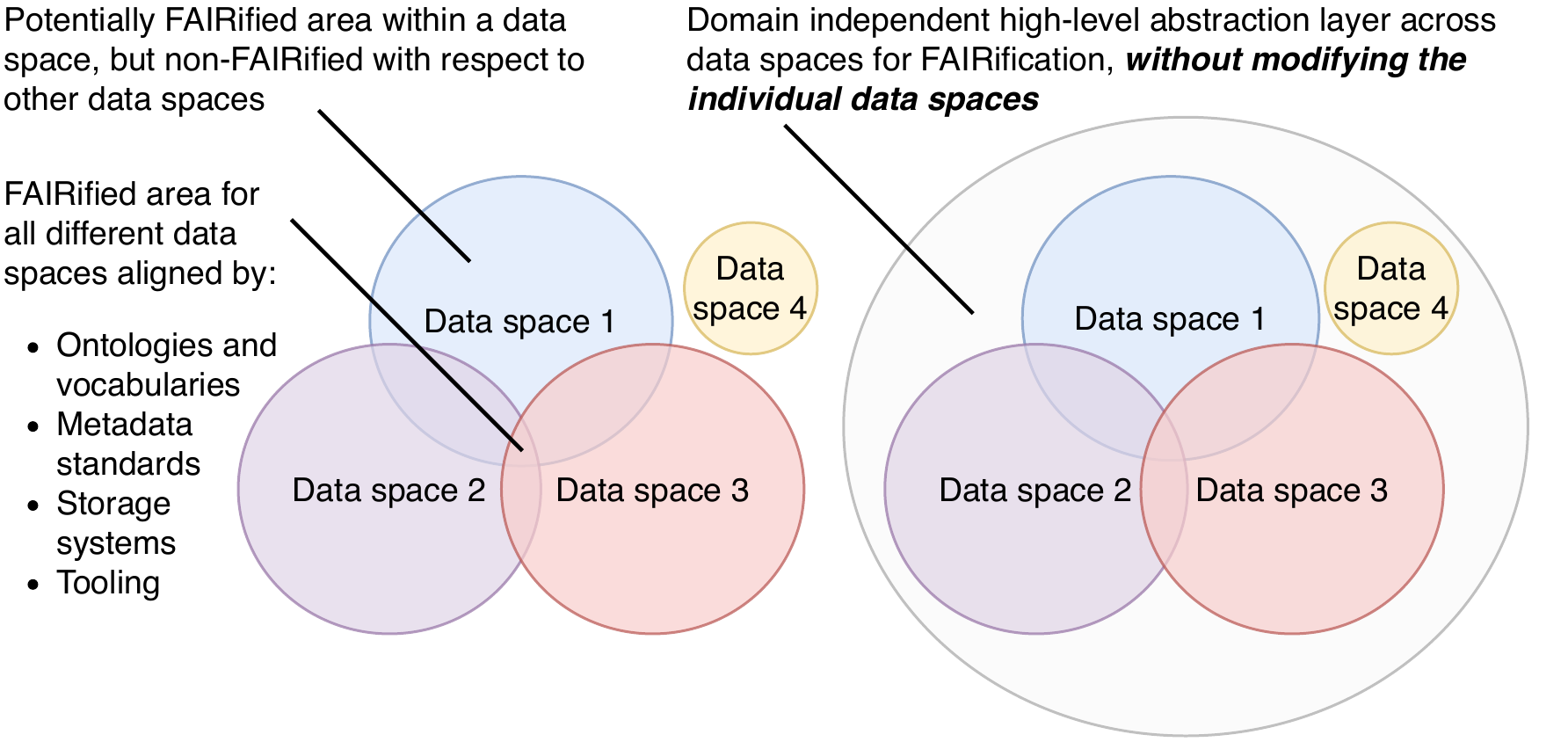}
    \caption{The current state of partially aligned FAIRified data spaces is illustrated by the Euler diagram on the left. Conversely, the diagram on the right includes an enclosed area that represents the abstraction layer which enables alignment across data spaces by providing an overarching FAIRified structure at the meta level.}
    \label{fig:dataspaceproblem}
\end{figure}

The advent of the FAIR principles for making data findable, accessible, interoperable and reusable provided a road map for achieving streamlined data management practices\cite{wilkinson_fair_2016}. "FAIRification" describes the process that organizations undergo to transform their data spaces into FAIR-compatible environments, a state that is dynamic and difficult to measure\cite{devaraju_automated_2021}. Whilst the FAIR principles provide guidelines for proper data management and stewardship, their implementation strategies have been realized in many ways\cite{jacobsen_fair_2020}. This is largely caused by the different requirements and practices within different communities. Therefore, different data spaces are typically only partially "FAIRified" in an interoperable way compared to each other, if at all, leading to less efficient research. A domain-independent, high-level abstraction layer could address this issue by providing a FAIR-compatible representation of each data space's contents without altering their native configurations, allowing each data space to maintain full control over its digital resources. This approach is illustrated in Figure~\ref{fig:dataspaceproblem}, where the overlapping regions in the Euler diagrams represent the FAIR-compliant elements shared among data spaces. However, these shared, interoperable regions are typically minimal or even non-existent, highlighting the challenges in achieving widespread FAIR compatibility across different domains.

FAIR Digital Objects (FDOs) focus on the reusability of individual data resources according to the FAIR principles and offer a lightweight approach to provide a uniform representation of diverse digital resources across data spaces, effectively abstracting from the user the complexity of handling their underlying bit sequences\cite{smedt_fair_2020, wittenburg_digital_2019}. Their conceptual approach at enabling machine-actionability provides particular perspectives for the enhancement of interoperability and reusability, which are the most difficult aspects of FAIR data to achieve \cite{wilkinson_interoperability_2017}. "Machine-actionable" in this context is a multifaceted term; in essence, it means that certain processes are performed automatically with little to no human intervention. This concept therefore constitutes one potential approach for realizing an abstracted high-level representation layer across different data spaces as previously described. However, a concrete formalization of the conceptual data model and interpretation of perspectives for aligning data spaces with respect to big data has not yet been introduced.

The remainder of this paper is structured as follows:
\begin{itemize}
    \item In section II, we describe the related works for FAIR principles and FDOs.
    \item Section III outlines a formal specification and FAIR criteria compliance analysis of an FDO data model that enables a standardized implementation of the FDO concept.
    \item Section IV provides implementation examples for two use cases and a comparative analysis of our model with FDO specifications used in other projects.
    \item Section V discusses the adoption requirements, limitations, and perspectives for big data research of our formalized FDO data model.
\end{itemize}

\section{Related Work}
\subsection{FAIR}
As originally formulated, the FAIR principles provide a set of best practices for managing research data and its metadata\cite{wilkinson_fair_2016}. Different communities adopted these principles and formulated additional requirements and aspects. For example, new FAIR requirements for research software (FAIR4RS) were considered separately by \cite{barker_introducing_2022}. Building on FAIR4RS, the unique characteristics of AI research—which involve a complex interplay of research data, metadata, and software for AI models—prompted the development of revised FAIR requirements tailored specifically for this field \cite{huerta_fair_2023, Katz_working_2021}. As stated by \cite{jacobsen_fair_2020}, the FAIR principles are formulated on a high level and allow for different interpretations and implementations, however ``for true interoperability we need to support convergence in implementation choices that are widely accessible and (re)-usable''.

An important aspect is thereby the provision of metadata for machines, i.e., structured data that enables automated systems to locate, interpret, and process digital resources reliably. The role of identifying this metadata and its typing is crucial to enable proper processing of the given information. This resulted in the specification of PID Information Types (PITs) which can be modeled hierarchically with a finite combination of PITs and Basic PITs down to the elementary level of JSON types for automated schema extraction as described in \cite{schwardmann_automated_2016}.
\subsection{FAIR Digital Objects}
\label{sec:FAIR Digital Objects}
The concept of Digital Objects was first introduced by \cite{kahn_framework_2006}, where a Digital Object is described as a data structure with associated components that can be queried using a globally available identifier system. A formal terminology definition has been formulated in a core data model for Digital Objects\footnote{\url{https://zenodo.org/records/2574407##.XG5s4OhKhaQ}} by the Data Foundation and Terminology working group of the Research Data Alliance (RDA)\footnote{\url{https://www.rd-alliance.org}}. The need for a transfer protocol for Digital Objects then led to the development of a uniform communication protocol called the Digital Object Interface Protocol (DOIP) \cite{doip_spec}. Building on this foundation, subsequent works have explored the relationship of Digital Objects to other concepts in IT, including the object-oriented programming (OOP) paradigm\cite{wittenburg_digital_2019}, the Semantic Web\cite{smedt_fair_2020}, the PID Graph\cite{cousijn_connected_2021}, and the services that make up an infrastructure for Digital Objects\cite{schwardmann_automated_2016, broeder_data_2014, schwardmann_digital_2020}.

The inherent characteristics of this concept were designed in a way that is compatible with the requirements of the FAIR principles and were addressed as a possible method for their implementation. This gave rise to the FAIR Digital Object (FDO) concept \cite{schultes_fair_2019}, with the potential to facilitate broader abstraction for interoperability in data management \cite{wittenburg_digital_2019}. This conceptual evolution is now being driven primarily by the RDA, the {European Open Science Cloud} (EOSC)\footnote{\url{https://eosc.eu}} as part of their interoperability framework\cite{doi/10.2777/620649}, CODATA\footnote{\url{https://codata.org}}, and the FDO Forum, which provides a list of specifications\footnote{\url{https://fairdo.org/specifications/}}. Several implementation examples resulted from these initiatives in the frame of different use cases such as in the domain of biodiversity\cite{deeleman-reinhold_genera_2024, islam_assessing_2023} or energy research\cite{mayer_thermal_2023}, whilst others modelled FDOs using specific methods and technologies, e.g. \cite{bonino_da_silva_santos_towards_2023} describes an OWL-based ontology for the FDO Framework. However, the different specifications of these communities and their approaches of implementing the FDO concept are restricted to (conceptual) semantic models, often lack in clarity, and partly contradict each other\cite{soiland-reyes_evaluating_2024}. This results in a similar problem of divergence as with the realization of FAIR principles by different communities. A use case-independent, mathematically rigorous, and formal data model is therefore required, enabling a standardized implementation and mechanism to validate what an FDO is, something that has not yet been sufficiently addressed. 

\section{FDO Data Model Formalization}
\begin{figure}[tbp]
    \centering
    \includegraphics[width=\columnwidth]{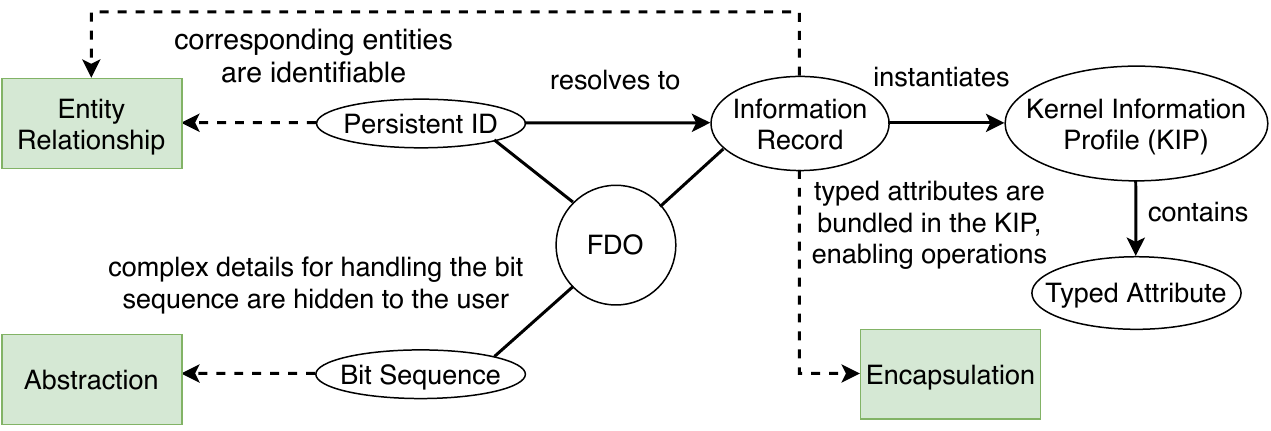}
    \caption{The semantic FDO Data Model specification depicting the relationships between FDO components and principles adopted from other fields of computer science, i.e., Abstraction, Encapsulation, and Entity Relationship.}
    \label{fig:fdo_data_model}
\end{figure}

In the following subsections, we formalize a data model for FDOs based on the original concept and the works developed by the aforementioned communities (cf. \cref{sec:FAIR Digital Objects}). As a basis for the formalization, we use the semantic data model illustrated in Figure~\ref{fig:fdo_data_model} that is detailed throughout this section. We define a set of formal expressions for key characteristics which supports the consistent adoption of the general concept while establishing a basis for validating whether an entity qualifies as an FDO and how it functions. We use the term \textit{digital resource} for any type of data in order to avoid confusion with the term \textit{digital object}. The model is based on the following initial case: one or more corresponding digital resources are stored in a permanent data storage system (possibly distributed), from where they are available as bit sequences. In the following sections, where we describe and formalize this model, we use the term \textit{typed} according to the definition of PITs and Basic PITs \cite{schwardmann_automated_2016} constituting the foundations for an FDO type system.

\subsection{Abstraction and Encapsulation}
\label{Abstraction and Encapsulation}
An FDO entity \(f\) in the set \( F \) of all FDOs requires an information record, denoted as a set \( R \), that instantiates an element from the set \( P \) of Kernel Information Profiles (KIPs). The concept of a KIP originates from the RDA\footnote{\url{https://zenodo.org/records/3581275}} and is described as a profile that contains and specifies a canonical set \( A_{p} \) of typed attributes describing the bit sequence represented by an FDO such that \( A_{p} \subseteq p \in P \). This enables syntactic validation by traversing the subtype hierarchies of information types, and potentially semantic validation when, for example, vocabulary terms are linked to the PID of these types. Providing the metadata for the essential kernel information about the bit sequence is required. We can then associate the creation of an FDO directly with the instantiation of a KIP:
\begin{equation}
\forall f \in F \ \exists p \in P: \text{Instantiate}(f,p)=R_{f}
\label{exp:exp1}
\end{equation}

By design, each FDO can only be associated with one information record \(R_{f}\) that instantiates exactly one KIP:
\begin{equation}
\begin{split}
    \forall f \in F: (\exists p \in P: \text{Instantiate}(f,p)&=R_{f} \land \\
     \exists p' \in P: \text{Instantiate}(f,p')&=R_{f}) \implies p=p'
\end{split}
\label{exp:exp2}
\end{equation}

The instantiation of a KIP requires that elements in a list \( V \) of valid non-empty values, characterizing the represented bit sequence, are successfully inserted for each attribute \( a_{p} \in A_{p} \) contained in the KIP according to its type, creating a key-value pair in the information record \( R_{f} \). According to the specification of \textit{typed} attributes in this context, the information record must expose the unambiguous PIDs of each attribute and not the human-readable names. A minimum set \( A_{m} \) of mandatory kernel information, as proposed by the RDA, comprises the following generic attributes, where \( A_{m} = \) \{\textit{Kernel Information Profile Reference, License, Checksum, Digital Resource Location, Creation Date, Digital Resource Type}\} \(\not=\varnothing\), such that \( A_{m} \subseteq A_{p}\). In addition, a Handle PID \( i \) in the set of PIDs \(I\) is assigned to the information record \(R_{f}\), creating the registered information record \( R^{f}_{reg} \) for a given FDO \(f\):
\begin{equation}
\begin{split}
\forall f \in F:\ (\forall a_{m}\in A_{m} \ \exists v\in V:\text{Create}(a_{m},v)&=\langle a_{m},v\rangle) \land \\
    \exists! i: \text{Assign}(i, R_{f})&=R^{f}_{reg}
\end{split}
\label{exp:exp3}
\end{equation}

The creation of optional key-value pairs \(\langle a_{p},v\rangle\), where \(a_{p} \in A_{p}\setminus A_{m}\), is not required for the successful creation of a valid FDO \(f\).

This formalization enables the representation of digital resources on the uniform FDO entity level with machine-interpretable characteristics. In this work, processes that operate on these entities are called \textit{operations}. In general, the operations that can be performed on FDOs, besides resolving the information record using the PID, are given as a set \( O \). The association between an FDO and its operations can be inferred by the set \(K \subset R\) of the typed attributes' (\(A_{m}\) and \(A_{p}\)) key-value pairs in the FDO information record \(R\) (now used equivalent to \(R^{f}_{reg}\)): 
\begin{equation}
\begin{split}
\forall f \in F: \ \exists k \in K \ \exists o \in O: \text{Associated}(k,\, o)&=O_{f}
\end{split}
\label{exp:exp4}
\end{equation}

The set \( O_{f} \) of operations that are associated with an FDO may be applicable to metadata \( m \) in the registered information record \( R \), e.g. checksum or creation date, or directly to the bit sequence \( s \) of the represented digital resource. \( O_{f} \) is thus decomposed into two subsets \( O_{m} \) and \( O_{s} \), with \( O_{m} \cap O_{s} \not=\varnothing\) in general. Given an attribute's key-value pair \( k_{acc}\) to access the bit sequence, e.g. the location reference, the following applies:
\begin{equation}
\forall o_{s} \in O_{s} \ \exists k_{acc}\in R : \text{Access}(k_{acc}) = s
\label{exp:exp5}
\end{equation}
The general applicability of associated operations based on a target \( t \) is then expressed as:
\begin{equation}
\forall o_{f} \in O_{f} \ \exists t \in \{m, s\} : \text{Applicable}(o_{f}, t) = \text{True}
\label{exp:exp6}
\end{equation}

Overall, these characteristics relate to the idea of an object, as understood in OOP, using encapsulation and abstraction. In this context, encapsulation means that the FDO bundles and exposes a set of typed attributes that are machine-interpretable and enable machine-actionable decisions that can be executed on the FDO. Abstraction is given by providing a uniform representation for handling the complex details of the data resource, which is available as a bit sequence.

\subsection{Entity Relationship}
\begin{figure}[tbp]
    \centering
    \includegraphics[width=\columnwidth]{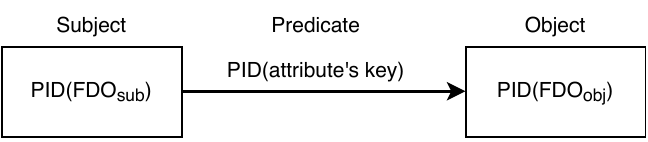}
    \caption{The conceptual model of PID triples based on the FDO's entity relationship characteristics in the spirit of RDF triples, connecting an FDO\textsubscript{sub} with an FDO\textsubscript{obj} by a typed attribute key working as predicate.}
    \label{fig: pid-triple}
\end{figure}

Given the set \( E \) of all entities, with \( F \subset E \), an FDO is linked to other FDOs or individual entities, e.g., on the internet, via a subset \( K' \subset K \) of the typed attributes' referencing key-value pairs in its information record, where the key specifies the relation to a target that is referenced by the value, which can be based on a PID or URI datatype:
\begin{equation}
\forall f \in F:\ K'_{f} \rightarrow E',\ k'_{f}\mapsto e'
\label{exp:exp7}
\end{equation}
This allows to determine how these entities are related to or may interact with each other. Referenced entities which are FDOs also provide useful information through their typed information record and may include links to other resources for further discovery. This relates to principles also used in Linked Data where RDF triples of URLs are used for interlinking data in the context of the Semantic Web. By the nature of FDOs for relating entities on the Web, including other FDOs, RDF triples can then be applied purely based on PIDs instead of URLs to constitute a directed graph that may also have strongly connected components and represents a given FDO space (refering to a conceptual environment for FDOs, rather than a space in the mathematical sense). In this case, the PID triple is given as \( \langle i_{sub}, i_{pre}, i_{obj}\rangle\), where \(i_{sub}\) is the PID of an FDO\textsubscript{sub} (the subject), \(i_{pre}\) is the PID of a typed attribute's key in the FDO\textsubscript{sub}'s information record that has a semantic definition (the predicate), and \(i_{obj}\) is the PID of an FDO\textsubscript{obj} (the object). This is illustrated in Figure~\ref{fig: pid-triple}. 

Denoting \( \tilde{F} \) as the set of PIDs with \( \tilde{F} \subset I\) that identify a set of FDOs, \( \tilde{K} \) as the set of all PIDs of typed attribute's keys that can act as predicates with \( \tilde{K} \subset I\), and \( T \) as the set of PID triples, the FDO graph structure \( G \) is then given as:
\begin{equation}
G \coloneq \left((\tilde{F}, \tilde{K}), T, g\right)
\label{exp:exp8}
\end{equation}
where \((\tilde{F}, \tilde{K})\) spans the graph of FDOs and \(g\) is defined as a function \(g: T \rightarrow \tilde{F} \times \tilde{K} \times \tilde{F},\,t\mapsto \langle i_{sub}, i_{pre}, i_{obj}\rangle\) mapping the triple \( t \) to an ordered triple of vertices connected by a predicated edge with the elements \(i_{sub}, i_{obj} \in \tilde{F}\), and \(i_{pre} \in \tilde{K}\).
\subsection{Fulfillment of FAIR Criteria}
To assess the FAIRification potential of this data model, we analyze its inherent fulfillment of the FAIR criteria. Thereby, we consider FAIR as originally formulated \cite{wilkinson_fair_2016}, as well as with a focus on research software (FAIR4RS) \cite{barker_introducing_2022}. Existing FAIR assessment tools like F-UJI\cite{devaraju_automated_2021} are currently not supporting the analysis of FDOs. Therefore, we describe here the characteristics of the FDO in relation to the FAIR criteria. FDOs represent digital resources on the individual entity level, e.g., scientific data, different types of metadata, such as experimental conditions, annotations, publications, or software. The granularity of this representation can be constantly increased, and related resources can be linked via this entity representation. This is the foundation for the following FAIR assessment:

\begin{itemize}
    \item Findability of all digital resources is independent of their storage system. Each resource is identifiable through a unique PID of their representing FDO that can be discovered, resolved, and interpreted. This enables the proper discovery and treatment of different digital resources according to the individual task. (cf. FAIR: F1 and FAIR4RS: F1-F1.2)
    \item FDOs can be related to other entities via their PID information records so that relationships between digital resources can be derived. (cf. FAIR: F3, I3 and FAIR4RS: F3, I2, R2)
    \item Each FDO is accessible by a communication protocol and actionable by a set of operations that are performed on the metadata in the information record and the bit sequence. This allows the digital resources represented by FDOs to be accessed, retrieved, and eventually manipulated without knowing the implementation details of the underlying technology. Thus, different digital resources represented by FDOs are actionable through a uniform interface. (cf. FAIR: A1-A1.2 and FAIR4RS: A1-A1.2, I1)
    \item FDO information records are persistently preserved through the policies of the PID system and independent of the existence of the digital resource they represent. This information may still serve for reproducibility of projects, although the involved digital resources are no longer available. (cf. FAIR: A2 and FAIR4RS: A2)
    \item The type system and consistent structure of FDOs allows for interoperable standards across communities as part of the existing FDO space. Thus, compatible resource types from different communities, such as datasets and applicable software, can be automatically identified and reused in the given context. (cf. FAIR: I1, I2 and FAIR4RS: I1, I2, R3)
    \item The information record contains a minimal set of relevant metadata attributes for each entity, defined by reusable KIPs and typed attributes. All types of digital resources can be described with a plurality of attributes required for their handling, e.g. licensing. By building these profiles and typed attributes using taxonomic and hierarchical structures, a common meaning of these contents and intended interpretation is facilitated for data reuse. Typed attributes aim towards the machine-interpretable description of existing standards and technologies that are associated and linked with the digital resource by the FDO's entity relationships. (cf. FAIR: F2, I1, I2, R1-R1.3 and FAIR4RS: R1-R3)
\end{itemize} 

\section{Implementation Examples}
In this section, we use the assertions given by the formal expressions of our formalized FDO data model for two exemplary cross-domain use cases. First, we elaborate on typical challenges that occur in the frame of these examples. We then describe the steps that can be performed using an FDO framework that leverages the assertions of our expressions to address these problems. Thereby, we provide a bilateral description, considering ingestion of FDOs in the FDO space that constitutes the high-level abstraction layer, as well as the retrieval in the context of data reuse based on an exemplary FDO. In the second phase, we perform a comparative analysis of our model compatibility with existing Handle records that were related to existing FDO specifications.

\subsection{Use Cases}
\paragraph{Energy Research}
Our first example comes from energy research. Datasets in this domain cover extensive, high-frequency, or high-resolution data across various energy sectors, enabling researchers to analyze patterns, improve efficiency, optimize energy systems, and forecast demand. The ``Thermal Bridges on Building Rooftops Dataset''\footnote{\url{https://doi.org/10.5281/zenodo.7022736}} is a representative, exemplary dataset that is used for identifying heat loss points on rooftops and could be valuable for energy conservation studies using AI methods as presented in \cite{mayer_deep_2023}. It is composed of several types of digital resources, comprising drone image sets, COCO annotation files, metadata files based on the Spatio Temporal Asset Catalogs (STAC), and Frictionless Data standards as described in \cite{mayer_thermal_2023}. 

A typical challenge regarding interoperability when working with this type of digital resource and AI applications is related to the tasks of identification and aggregation of other datasets. These may originate from a different data space (e.g. weather data) using metadata, data integration, transformation, preprocessing, analytics and integrity checks across systems and data spaces, operating with various technologies and tools.
\\
\paragraph{Digital Humanities}
The second use case considers the domain of digital humanities (DH) where ontologies, thesauri and controlled vocabularies (CVs) are leveraged that consist of various terminologies, languages, and data models (e.g. the Simple Knowledge Organization System (SKOS)) for analyzing cultural, historical, and linguistic phenomena. 
A typical task in this field is related to ontology matching, which helps in aligning and integrating heterogeneous ontologies, thesauri, and CVs.
The "Ontology Matching Benchmark Dataset for Digital Humanities" \cite{krausGoldStandardBenchmark2024} is an example for a multilingual and SKOS-based dataset which is composed of multiple CVs and applicable for the task of advancing matching system development when more CVs are added.

In contrast to our first example, we do not only consider the applicability of this dataset, but the prior challenge of composing it from various CVs with respect to interoperability. This typically comprises the tasks of domain identification, CV quantification, conversion, SKOS and format validation, as well as aggregation and integration.

\subsection{FDO framework}
\label{subsec:fdo framework}
We now further describe the creation of FDOs based on a concrete example with respect to our data model, followed by its (re)use in the context of the use cases. As previously defined \cref{Abstraction and Encapsulation}, in order to create FDOs for the given digital resources, a KIP has to be defined and instantiated that conforms to the requirements of the mandatory kernel information set \(A_{m}\). For this, we choose the \textit{Helmholtz KIP} \cite{jejkal_realizing_2022} that corresponds to the recommendations of the RDA. This profile contains, besides others, the typed attributes shown in Table \ref{table:metadata_fields} which were used in our examples. Attribute PIDs and additional type specifications such as regular expressions are left out for space reasons. For this, we refer to the specifications in the ePIC Information Type Registry\footnote{\url{https://dtr.pidconsortium.eu}} where a version of this KIP was registered and can be resolved using the Handle Registry proxy server\footnote{\url{https://hdl.handle.net}} using the PID \href{https://hdl.handle.net/21.T11148/b9b76f887845e32d29f7}{21.T11148/b9b76f887845e32d29f7}.

\begin{table}[tbp]
\caption{Typed attributes of the Helmholtz KIP.}
\centering
\scriptsize
\begin{tabular}{|p{2.4cm}|p{0.6cm}|p{0.8cm}|p{3.4cm}|}
\hline
\textbf{Name} & \textbf{Obli-\linebreak gatory} & \textbf{Repea-\linebreak table} & \textbf{Value Type} \\
\hline
kernelInformationProfile & yes & no & Handle-Identifier-ASCII \\
digitalResourceLocation & yes & yes & URL \\
dateCreated & yes & no & date-time-rfc3339 \\
dateModified & no & no & date-time-rfc3339 \\
license & yes & no & URL \\
digitalResourceType & yes & no & media-type-IANA \\
checksum & yes & no & checksum-string \\
version & no & no & version-number \\
hasMetadata & no & yes & Handle-Identifier-ASCII \\
isMetadataFor & no & yes & Handle-Identifier-ASCII \\
hasSchema & no & no & URL \\
topic & no & yes & URL \\
contact & no & yes & URL \\
identifier & no & yes & string \\
DataCite-Language & no & yes & Language-Codes-ISO-639-1 \\
\hline
\end{tabular}
\label{table:metadata_fields}
\end{table}

\begin{figure}[tbp]
    \centering
    \includegraphics[width=\columnwidth]{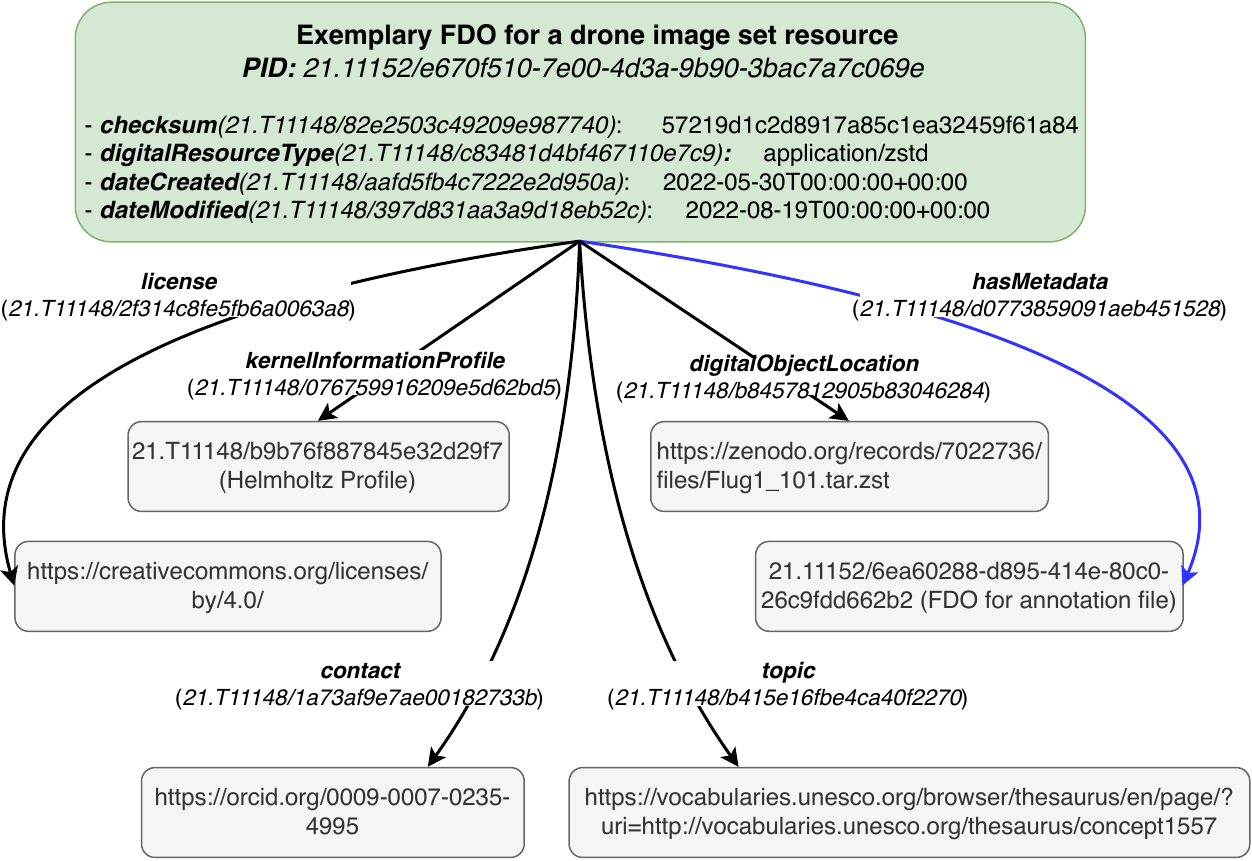}
    \caption{An exemplary FDO according to the formalized data model that contains a set of non-referencing and referencing typed attributes in its information record. The latter enables entity relationships, including FDO-FDO relations (pointed out with a blue arrow) by PID-triples.}
    \label{fig:example_fdo}
\end{figure}

It can be noticed that the set of typed attributes in  this profile contains elements beyond the mandatory attributes, which is consistent with our model specification. However, which particular additional attributes should be contained in a profile and therefore present in the information record, i.e., which kernel information should be available on the FDO level, as well as the rules for extending these profiles, is not addressed in the frame of this work. This has to be defined with a set of rigorous rules in compliance with the FDO type system to ensure interoperability. As in the case of the mandatory kernel information, these attributes should primarily serve the purpose of associating useful operations to the FDO and relate to other entities.
Applying \eqref{exp:exp1} to \eqref{exp:exp3}, this KIP was instantiated using an instance of the Typed PID Maker\footnote{\url{https://github.com/kit-data-manager/pit-service}}, which is a prototype for a PIT service, yielding an information record \(R^{f}_{reg}\) (cf. \eqref{exp:exp3}) that is registered with the Handle Registry. According to \eqref{exp:exp7}, the entity relationships are given by those attribute key-values pairs that have a reference type as value. This also comprises the PID-triples for the resulting FDO graph \(G\). An exemplary FDO picked from the energy research use case data is illustrated in Figure~\ref{fig:example_fdo}. In the following, we show the values for the nodes \(\tilde{F}\), predicated edges \(\tilde{K}\), and triples \(T\) for a corresponding directed FDO graph \(G\) as given by \eqref{exp:exp8} that covers a minimal network for this set of FDOs where each has at least one connection to another entity: 
\begin{equation*}
\begin{split}
&\tilde{F} = \{ A, B, C, D, E, F, G, H, I, J, K, L, M, N, O, P, Q, R \}\\
&\tilde{K} = \{ a, b\}\\
&T = \{ ( A, a, B ), ( C, b, H), ( D, a, K), ( E, a, M ), ( F, a, I ),\\
&( G, a, O ), (J, b, L), (N, b, F), (P, a, J), (Q, a, J), (R, a, J)\}
\end{split}
\end{equation*}
where:
\begin{itemize}
    \item[] A = \href{https://hdl.handle.net/21.11152/e670f510-7e00-4d3a-9b90-3bac7a7c069e}{drone image set 1}
    \item[] B = \href{https://hdl.handle.net/21.11152/6ea60288-d895-414e-80c0-26c9fdd662b2}{annotation file 1}
    \item[] C = \href{https://hdl.handle.net/21.11152/58d43ddc-5e29-4980-8675-ae579b50a1e2}{annotation file 2}
    \item[] D = \href{https://hdl.handle.net/21.11152/6858a0b5-cc60-40e9-afef-8c2dd8b35e8e}{drone image set 2}
    \item[] E = \href{https://hdl.handle.net/21.11152/3ab9f444-05f6-445e-a691-62fae4021bea}{drone image set 3}
    \item[] F = \href{https://hdl.handle.net/21.11152/365fd8cf-8e86-41b8-9d0e-b816fdd01d29}{drone image set 4}
    \item[] G = \href{https://hdl.handle.net/21.11152/041a6111-644a-4617-afb3-3c421a88e8e3}{drone image set 5}
    \item[] H = \href{https://hdl.handle.net/21.11152/f48bf4e7-3879-4216-8f64-45a060b8f658}{drone image set 6}
    \item[] I = \href{https://hdl.handle.net/21.11152/7b58b3b5-75eb-4417-ac4d-abe025e159f6}{frictionless data standard file}
    \item[] J = \href{https://hdl.handle.net/21.11152/ba370aa3-6422-428c-9ff7-c2ef429df603}{STAC collection file}
    \item[] K = \href{https://hdl.handle.net/21.11152/09cb76fc-b8cb-4116-a22a-68c5bdfa77b0}{STAC feature file 1}
    \item[] L = \href{https://hdl.handle.net/21.11152/24a55398-b96b-43dd-b0fb-cd8ce302c7ce}{STAC feature file 2}
    \item[] M = \href{https://hdl.handle.net/21.11152/721234ac-4b5a-4d02-9944-82a08ef2db35}{STAC feature file 3}
    \item[] N = \href{https://hdl.handle.net/21.11152/ebaeb5bc-0514-47c9-bcd2-98f0253843d8}{STAC feature file 4}
    \item[] O = \href{https://hdl.handle.net/21.11152/9854677c-77c5-4a0b-916b-57dd9ec20198}{STAC feature file 5}
    \item[] P = \href{https://hdl.handle.net/21.11152/cfd0fc0e-f5ea-464e-a57f-28e882924860}{STAC feature file 6}
    \item[] Q = \href{https://hdl.handle.net/21.11152/976fcf28-f924-4a21-b53d-5d054ad8198d}{STAC camera file 1}
    \item[] R = \href{https://hdl.handle.net/21.11152/37833c54-1d36-42e4-858d-831447122863}{STAC camera file 2}
\\
    \item[] a = \href{https://hdl.handle.net/21.T11148/d0773859091aeb451528}{hasMetadata}
    \item[] b = \href{https://hdl.handle.net/21.T11148/4fe7cde52629b61e3b82}{isMetadataFor}
\end{itemize}
All of these FDOs can be resolved by their PIDs (cf. referenced Zenodo repository) via the Handle Registry in order to reproduce the entire FDO graph with all connections, also revealing strongly-connected components.  

Having all digital resources represented in this uniform structure, \eqref{exp:exp4} can be used to yield the set \(O_{f}\) of operations for each FDO to tackle the challenges described earlier for each use case example by enabling a machine-actionable decision
framework that abstracts the details of domain and technology
specific information to the client. Regarding these tasks that must be performed, we define a set of example operations that are associated with FDOs via one or more attribute key-value pairs in their information records. According to \eqref{exp:exp5} and \eqref{exp:exp6}, these operations are applicable to a respective target, i.e., the metadata in the information record or the bit sequence. Whilst some operations will be naturally very specific to a particular type of digital resource, typically when being applied to the bit sequence, others are more generic. For example, they could process the kernel information to:
\begin{itemize}
    \item evaluate reusability of the digital resource based on the license information.
    \item validate integrity of the resource based on the checksum value.
    \item traverse the FDO graph structure given by the entity relationships based on PID references.
    \item retrieve a digital resource from its storage system which is contained in a particular data space and is crucial for any following operation on the bit sequence according to \eqref{exp:exp5}.
\end{itemize}
These operations can be directly associated with a subset of the typed attributes in the information records of the shown example FDOs. A subset of operations for all FDOs in this example may therefore be \textsc{evaluate\_license(), validate\_checksum(), get\_related\_fdo(), get\_digital\_resource()} with the respective machine-interpretable typed attribute's key-value subsets (only considering the key) \textit{\{license\}, \{checksum\}, \{hasMetadata, isMetadataFor\}, \{digitalResourceLocation\}} being the association criteria. By the underlying type system of FDOs as per our data model, all digital resources from any data space that are represented on the FDO level would have the same set of such generic operations to make machine-actionable decisions based on FDO entities.
In the following, we will show how the same principles can be applied to the more specific problems of our use cases, outlining how FDOs can help in solving the challenges mentioned earlier.
\\
In order to preselect drone images, different metadata can be utilized such as geographic regions or time frames which are typically included in specific schemas such as STAC. The STAC files associated with a given drone image set can be discovered by the FDO entity relationships and a corresponding operation like \textsc{get\_related\_fdo()} as indicated earlier. An operation for filtering tasks on these resources could be called \textsc{geographic\_filter()} and \textsc{timestamp\_filter()}. These would have to be applied directly to the bit sequence that contains the geographic metadata (latitude/longitude, region codes) for isolating data from specific areas, or timestamp metadata for filtering images or measurements from specific periods. The association criteria for such operations could therefore be related to the typed attribute's key-value subset (considering either only the key or the key-value pair) of \textit{\{(hasSchema, v\textsubscript{1}), (digitalResourceType, v\textsubscript{2}), digitalResourceLocation\}}, where \textit{v\textsubscript{1}} could be a value indicating the STAC specification such as ``https://schemas.stacspec.org/\textellipsis'', and \textit{v\textsubscript{2}} the value of the data type for this bit sequence given as MIME type "application/json".

It can be observed that these operations can be employed to access the domain-specific metadata based on the kernel information of the FDO, and to process them in a subsequent step. The sequence of steps and the actual processing done by these operations is thereby abstracted for the user by the entity relationship and encapsulation characteristics of the FDO. Likewise, additional operations may be utilized to identify and aggregate distinct but relevant datasets from different data spaces by applying the required operations to traverse the FDO space and process their kernel information and bit sequences dependent on the given task to get to the existing domain-specific information. Especially regarding enhanced semantic interoperability, such operations could highly profit from well established Semantic Web and Linked Data principles. Further automated preprocessing, such as image rescaling and normalization, pairing with annotation files or transformation of time-series data prior to AI applications, is a plausible use-case example for using FDOs in a big data context, but requires further clarification of the operation mechanisms. 
Likewise, the example of vocabularies for digital humanities becomes a big data challenge when entire digitized libraries spanning centuries, or texts from different languages, time periods, and genres are used in AI approaches like ontology-matching or text mining using NLP to identify patterns across centuries.
At this point, we don't further consider theoretical details of such a use case or relate it to the previous scenario from energy research, where digital resources from different data spaces are processed by an FDO framework. Instead, we again emphasize the data type- and technology-agnostic modelling approach provided by FDOs. Based on the \href{https://hdl.handle.net/21.11152/9f22289c-695e-476e-81b3-f806f6346654}{handle record} of an FDO representing a vocabulary entry from the aforementioned use case it can be seen that for a digital resource coming from a different data space, the FDO representation is still consistent (cf. \href{https://hdl.handle.net/21.11152/e670f510-7e00-4d3a-9b90-3bac7a7c069e}{drone image set}) and corresponds to the previously described FDO type system that uses PITs.

\begin{figure}[tbp]
    \centering
    \includegraphics[width=\columnwidth]{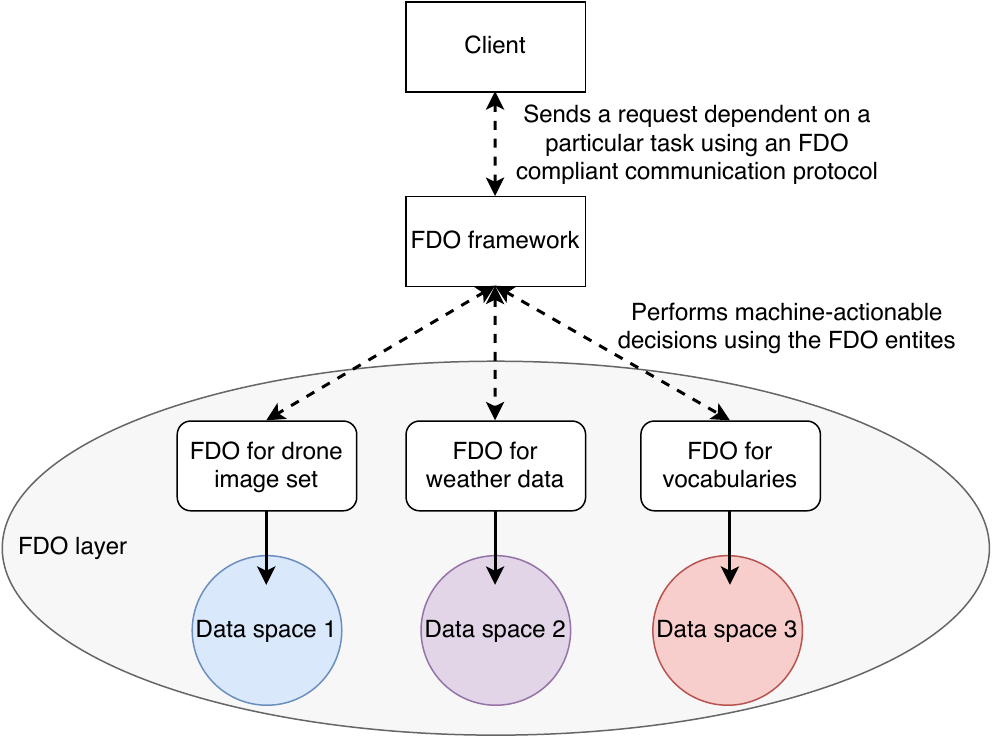}
    \caption{Resuming the illustration of a high-level abstraction layer around individual data spaces, FDOs can be considered retrievable and operable objects within this layer, representing digital resources within data spaces. A framework that uses their type system, operations, and entity linkage finally enables interoperability between the data spaces the FDOs point to.}
    \label{fig:fdo_framework}
\end{figure}

Independent of any particular use case, the aim of FDOs is to manage and process these digital resources using a minimal high-level abstraction layer, providing a framework to leverage existing technologies and tools across data spaces in an interoperable way, without the client needing to know about the details as illustrated in Figure~\ref{fig:fdo_framework}. This also requires a suitable data structure for specifying and calling associated operations. However, the exact implementation of operations, how their association criteria are inferred based on the syntactics and semantics of the typed attributes, and how they are finally executed on demand by client requests using a communication protocol, also in the frame of workflows, is out of scope for this paper. It is in any case important that an FDO framework has a set of known rules to operate on the given type system that should be standardized within an FDO space.
\subsection{Comparative Analysis}

As described in the state-of-the-art analysis (cf. \ref{sec:FAIR Digital Objects}), there currently exists a lack in uniformity for the modelling of FDOs by different communities, imposing a problem with respect to interoperability on the FDO level. In the following, we pick a set of existing Handle PID records that were related to the concept of FDOs in earlier works and validate their compatibility with our FDO data model. As a baseline, we use the FDO that we have shown in the earlier example, which fully corresponds to our model specification. Table \ref{table:handle-records-comparison} summarizes this comparison regarding the expressions that we defined for our data model, stating if they are fulfilled and giving additional explanation where required. We thereby don't consider the match of particular PIDs of the individual typed attributes which are currently not yet standardized. 
In the following, we will briefly describe the contexts on how these Handle PIDs are related to FDOs.
Overall, these results show that FDOs based on their current specifications may have some characteristics in common but are not fully compatible. This is challenging when attempting to provide a uniform layer for operating on the digital resources, represented by FDOs in an interoperable way.

\begin{table}[tbp]
\caption{The summary of the comparison between multiple Handle Records that are specified as FDOs with respect to the formalized FDO data model expressions.}
\centering
\scriptsize
\begin{tabular}{|p{3cm}|p{1.5cm}|p{1.2cm}|p{1.5cm}|}
\hline
\textbf{Context} & \textbf{PIDINST} & \textbf{DARIAH} & \textbf{DiSSCo} \\
\hline
Handle PID & 21.T11998/\linebreak 0000-001A-3905-1 & 21.11113/\linebreak 0000-000B-CA4C-D & 10.3535/\linebreak G0G-G7D-N5J \\
\hline
Instantiates one KIP (exp. 1, 2) & \textbf{yes} & \textbf{no} & \textbf{yes} \\
\hline
Attributes are typed on the record level can be validated (exp. 3) and associated with operations (exp. 4) & \textbf{partially} - not the KIP reference attribute & \textbf{no} & \textbf{partially} - attributes are not identifiable by their PID in the record \\
\hline
Contains the set of mandatory typed attributes, i.e. 6 (exp. 3) & \textbf{no} & \textbf{no} & \textbf{no} \\
\hline
Operations can access and subsequently process the bit sequence of the digital resource (exp. 5, 6) & \textbf{no} - provides a landing page as digital resource location & \textbf{no} & \textbf{no} - digital resource location is missing \\
\hline
Relates to other entities including other FDOs by PID-triples via typed attributes (exp. 7, 8) & \textbf{partially} - relates to other entities via URLs but not to other FDOs & \textbf{no} & \textbf{partially} - relates to other entities via URLs but not to other FDOs \\
\hline
\end{tabular}
\label{table:handle-records-comparison}
\end{table}

\paragraph{PIDINST}
Originating from the Persistent Identification of the Instrument (PIDINST) schema to reference and describe scientific instruments\cite{Stocker-2020}, the example Handle Record resolvable with the PID \href{https://hdl.handle.net/21.T11998/0000-001A-3905-1?noredirect}{21.T11998/0000-001A-3905-1} was described as almost fully compliant with the FDO specifications, i.e., the FDO information record, by \cite{schwardmann_how_2023}. 
This FDO example complies to a fair degree with the model requirements, especially regarding the type system, i.e., it contains to a large extent typed attributes, although it misses the majority of mandatory attributes we defined in \cref{Abstraction and Encapsulation}. Missing a typed KIP reference is problematic for validation, and the provision of a non-machine readable landing page as resource reference massively impedes the application of operations on the bit sequence level. 
\paragraph{DARIAH}
In the context of the German DARIAH project\footnote{\url{https://www.dariah.eu}} within the domain of digital humanities, the following Handle Record resolvable with the PID \href{https://hdl.handle.net/21.11113/0000-000B-CA4C-D?noredirect}{21.11113/0000-000B-CA4C-D} was used as an example for an FDO profile of "Legacy Repository Records" in \cite{schwardmann_how_2023}, stating that it is almost fully compliant with the FDO information record requirements.
Although the information record elements of this example FDO were retrospectively modeled in a type registry, their types cannot be easily assessed since no KIP was used in the first place to instantiate the record and therefore are not inferable. Consequently, all other requirements are also not met, which results from the earlier mentioned missing requirements.
\paragraph{DiSSCo}
As part of the DiSSCo project\footnote{\url{https://www.dissco.eu}}, the following Handle Record resolvable with the PID \href{https://hdl.handle.net/10.3535/G0G-G7D-N5J?noredirect}{10.3535/G0G-G7D-N5J} is used as FDO to represent digital specimen as described in \cite{deeleman-reinhold_genera_2024}.
Whilst this Handle record is compatible with our model specification in some aspects, it lacks the provision of PIDs for the registered typed attributes, which impairs unambiguous typing for machine-interpretability and subsequent actionability by the association of operations. It also does not comply with the minimum attribute set to access and process the bit sequence, e.g., the digital resource location, we require in our model. This is primarily because this record is a kind of self-describing representation of a non-digital entity of which it constitutes the bit-sequence, similar to the concept of a digital twin. Therefore, it also contains a relatively large amount of additional attributes that describe this specimen of which several are indeed consistent with the FDO attempt for providing kernel information and outline how an extended FDO record could look like. However, this does not comply directly with the data model we provide where an FDO is considered a representation of an existing, curated digital resource. Furthermore, only a few entity relationships are provided, of which some reference non-machine-readable landing pages.

\section{Discussion of Results}
\subsection{Adoption Requirements and Limitations}

\textit{Practical aspects.} The formalized FDO data model provides a baseline for aligned implementations and makes certain assumptions for adoption by communities that want to represent the independently existing and unmodified contents of their data spaces on this unified layer. FDOs are designed to represent and describe digital resources as uniform and abstracted high-level information in a way to make these curated and persistent resources machine-actionable. Digital resources that complement each other or can be commonly used in a given context are thereby connected on a meta-level. This makes FDOs foremost suitable to be used as a starting point in the initial phase of a research task to navigate through data spaces that are inherently not fully aligned in aspects of FAIR and to evaluate the usability of their digital resources. In advanced applications, the encapsulated operations associated with the FDO entities could also enable the execution of more complex workflows on the bit sequence level, gradually decreasing the overhead for the user. This could finally result in an ecosystem where a digital resource gets activated by an initial client request and performs a sequence of steps based on its given set of rules by the type system to deliver a particular aspect, modified version of itself, or a new result back to the client.
\newline

\textit{Organizational aspects.} Each community must decide which digital resources they want to represent according to the FDO specification and thus share on this abstracted layer. Central to this is the transfer of required metadata to the information record that is standardized by controlled and registered Kernel Information Profiles containing typed attributes. Defining these components is a process ideally performed on a global basis, analogue to RFCs, where initiatives such as the RDA or the FDO Forum provide exchange platforms. Based on generic types, communities may then extend and align their kernel information requirements consistently, whilst sticking to the agreed basis.

With respect to the infrastructure, we assume the use of certain base services is central, such as a PID service, e.g. the Handle Registry. In addition, KIPs and typed attributes for the information record can be reliably managed by possibly federated and aligned Data Type Registries. One example of such a registry is described in \cite{schwardmann_automated_2016}. The actual transfer, management of, and interaction with FDOs can be individually realized by each community that has shared access to these base services. FDO validation according to our formalized data model, i.e., using assertions \eqref{exp:exp1} to \eqref{exp:exp3}, must thereby be considered. A reference implementation for how this could be realized already exists as part of the Typed PID Maker (cf. \cref{subsec:fdo framework}).
\newline

\textit{Technical aspects.}
With respect to a communication protocol for interacting with FDOs, the current state-of-the-art internet protocol HTTP is suitable, but a more specific protocol is under discussion as described in \cite{kahn_framework_2006, doip_spec}. This so-called Digital Object Interface Protocol (DOIP) can be implemented directly via TCP/IP or as DOIP over HTTP. Technically, the idea is to trigger the utilization of the typed and machine-actionable FDO information record by requesting the entity's PID. With this protocol, the FDO characteristics of long-term preservation, enhanced interoperability and facilitated automation using operations are emphasized. Whilst this protocol is complementary with existing standards, it requires the additional adoption of syntactic rules for making PID requests. 

In order to implement the association between FDOs and applicable operations (\eqref{exp:exp4} to \eqref{exp:exp6}), a mechanism is required, which can be either static or dynamic. As long as such a mechanism utilizes the typed attributes in the FDO's information record, it can be expected to be compatible with our data model. Currently, typed attributes enable a reliable syntactic interpretation but are limited in aspects of dynamic semantic interpretation. Leveraging existing semantic methods and technologies for these types, e.g. controlled vocabularies, could enhance their capabilities in these regards. This is an organizational aspect as well. Abstract FDO entity relationships and FDO networks according to \eqref{exp:exp7} and  \eqref{exp:exp8} must be utilized with proper data structures, e.g. graphs. This would enable enhanced query performance and application of graph methods such as traversal algorithms, path finding or clustering. Additional components of the FDO data model could also be incorporated, such as the PITs of KIPs or typed attributes to reveal and utilize connections between isolated areas that originate from separate data spaces. With respect to the aspect of findability of entities, an FDO graph could also be semantically enriched and searched based on Semantic Web technologies. This would constitute a variant of a Knowledge Graph that represents the FDO space which is finally machine-actionable by an underlying type system conforming to our model. 

The formalization currently does not consider any security aspects which are, however, a topic of relevance on the conceptual level of FDOs. In doing so, data owners may decide to share only certain non-sensitive information of a digital resource and provide a standardized mechanism for authorized clients to access and operate on private contents. A typical use case would be in the realm of clinical data.

\subsection{Perspectives for Big Data Research}
Based on our FDO data model implementation and the results of our experiments, we see perspectives for this approach to enhance the usability and value of big data across various scientific and research domains, facilitating more effective and efficient data-driven discoveries.

The characteristics of the 5 Vs for describing big data typically take effect across data spaces, for which various technologies exist to take them into account. Here, centralized and distributed storage systems like repositories, HDFS, cloud storage, or data lakes are used to deal with large volumes. There also exists extensive tooling for data integration, wrangling, and quality assessment. As we have described in the introduction, the alignment of data spaces by FAIR principles is impeded by differences in their interpretation and implementation. FDOs, as per concept, aim towards the alignment of data spaces on a higher abstraction level, overcoming these heterogeneities. Implementing FDOs using a rigorously defined data model, as presented, is crucial in these regards. With respect to the 5 V dimensions, the benefits that are introduced by FDOs that adhere to a formalized data model can be summarized as follows:
\begin{itemize}
    \item \textit{Volume}: Large amounts of distributed or corresponding digital resources are linked by machine-actionable entities that can be discovered, accessed and analyzed using PID information records based on a type system in adherence to their kernel information.
    \item \textit{Variety}: Typed information records abstract the variety of the underlying bit sequences, facilitating the integration and aggregation of diverse digital resources.
    \item \textit{Velocity}: Entities described with typed information records allow assumptions to be made about how to operate on these entities, which can then be analyzed more quickly by automated processes.
    \item \textit{Veracity}: Comprehensive metadata are encapsulated in the typed information record such as checksums that can be automatically evaluated, verifying data reliability.
    \item \textit{Value}: The assessment of potential use cases for the underlying bit sequence using metadata is facilitated.
\end{itemize}

The resulting potential solutions that we have listed here primarily support the initial discovery phase, when large amounts of data are aggregated and evaluated for a specific use case, for example when considering the application of data in an AI project. The existing and diverse technologies are still crucial for managing these data resources, whilst FDOs could facilitate their handling. Their machine-actionable character paves the way towards a highly automated big data environment that requires only minor manual intervention by humans. FDOs have therefore the potential to align data spaces on a global scale without changing the running systems. 

\section{Conclusions}
FAIRification is a way to achieve globally aligned data spaces, enabling more enhanced utilization of their digital resources. FAIR Digital Objects can serve as a potential solution to the current limitations in this regard, without changing the individual data spaces. 
This work presents an attempt to formalize and standardize a model for the implementation of FDOs. Our analysis highlights the model's compliance with the FAIR principle criteria, with a particular focus on machine actionability. This formalism provides a baseline for communities attempting to share a high-level abstraction layer for representations of their digital resources. The strategy is to not re-invent the wheel, but to provide a lightweight solution (considering the cost-benefit ratio) that can be put on top of already established technologies and standards. The robustness of this FDO data model needs to be tested in future works by applying more complex use cases and workflows. Big data applications could highly profit from that with respect to the 5 Vs. The tooling needs to be further developed to ensure a wide adoption of FDOs in multiple research groups and domains. We believe that consensus on this meta-level is the key to a long-lasting and widely used layer for aligned FAIR data spaces. Our work provides the baseline for this.

\bibliographystyle{IEEEtran}
\bibliography{IEEEabrv,bstcontrol,mybibliography}

\begin{thebibliography}{10}
\providecommand{\url}[1]{#1}
\csname url@samestyle\endcsname
\providecommand{\newblock}{\relax}
\providecommand{\bibinfo}[2]{#2}
\providecommand{\BIBentrySTDinterwordspacing}{\spaceskip=0pt\relax}
\providecommand{\BIBentryALTinterwordstretchfactor}{4}
\providecommand{\BIBentryALTinterwordspacing}{\spaceskip=\fontdimen2\font plus
\BIBentryALTinterwordstretchfactor\fontdimen3\font minus \fontdimen4\font\relax}
\providecommand{\BIBforeignlanguage}[2]{{%
\expandafter\ifx\csname l@#1\endcsname\relax
\typeout{** WARNING: IEEEtran.bst: No hyphenation pattern has been}%
\typeout{** loaded for the language `#1'. Using the pattern for}%
\typeout{** the default language instead.}%
\else
\language=\csname l@#1\endcsname
\fi
#2}}
\providecommand{\BIBdecl}{\relax}
\BIBdecl

\bibitem{wilkinson_fair_2016}
M.~D. Wilkinson, M.~Dumontier, I.~J. Aalbersberg, G.~Appleton, M.~Axton, A.~Baak \emph{et~al.}, ``The {FAIR} {Guiding} {Principles} for scientific data management and stewardship,'' \emph{Sci. Data}, vol.~3, no.~1, p. 160018, 2016.

\bibitem{wittenburg_digital_2019}
\BIBentryALTinterwordspacing
P.~Wittenburg and G.~O. Strawn, ``Digital {Objects} as {Drivers} towards {Convergence} in {Data} {Infrastructures},'' 2019. [Online]. Available: \url{https://doi.org/10.23728/B2SHARE.B605D85809CA45679B110719B6C6CB11}
\BIBentrySTDinterwordspacing

\bibitem{wilkinson_interoperability_2017}
M.~D. Wilkinson, R.~Verborgh, L.~O. B. d.~S. Santos, T.~Clark, M.~A. Swertz, F.~D.~L. Kelpin \emph{et~al.}, ``\BIBforeignlanguage{en}{Interoperability and {FAIRness} through a novel combination of {Web} technologies},'' \emph{\BIBforeignlanguage{en}{PeerJ Comput. Sci.}}, vol.~3, p. e110, 2017.

\bibitem{curry_data_2022}
E.~Curry, S.~Scerri, and T.~Tuikka, \emph{Data Spaces: Design, Deployment, and Future Directions}.\hskip 1em plus 0.5em minus 0.4em\relax Cham: Springer International Publishing, 2022, pp. 1--17.

\bibitem{delgado_interoperability_2021}
J.~Delgado, \emph{Interoperability Effect in Big Data}.\hskip 1em plus 0.5em minus 0.4em\relax Cham: Springer International Publishing, 2021, pp. 875--901.

\bibitem{huerta_fair_2023}
E.~A. Huerta, B.~Blaiszik, L.~C. Brinson, K.~E. Bouchard, D.~Diaz, C.~Doglioni \emph{et~al.}, ``\BIBforeignlanguage{en}{{FAIR} for {AI}: {An} interdisciplinary and international community building perspective},'' \emph{\BIBforeignlanguage{en}{Sci. Data}}, vol.~10, no.~1, p. 487, Jul. 2023.

\bibitem{devaraju_automated_2021}
A.~Devaraju and R.~Huber, ``An automated solution for measuring the progress toward {FAIR} research data,'' \emph{Patterns}, vol.~2, no.~11, p. 100370, 2021.

\bibitem{jacobsen_fair_2020}
A.~Jacobsen, R.~de~Miranda~Azevedo, N.~Juty, D.~Batista, S.~Coles, R.~Cornet \emph{et~al.}, ``{FAIR} {Principles}: {Interpretations} and {Implementation} {Considerations},'' \emph{Data Intell.}, vol.~2, no. 1-2, pp. 10--29, Jan. 2020.

\bibitem{smedt_fair_2020}
K.~Smedt, D.~Koureas, and P.~Wittenburg, ``{FAIR} {Digital} {Objects} for {Science}: {From} {Data} {Pieces} to {Actionable} {Knowledge} {Units},'' \emph{Publications}, vol.~8, p.~21, 2020.

\bibitem{barker_introducing_2022}
M.~Barker, N.~P. Chue~Hong, D.~S. Katz, A.-L. Lamprecht, C.~Martinez-Ortiz, F.~Psomopoulos \emph{et~al.}, ``\BIBforeignlanguage{en}{Introducing the {FAIR} {Principles} for research software},'' \emph{\BIBforeignlanguage{en}{Sci. Data}}, vol.~9, no.~1, p. 622, Oct. 2022.

\bibitem{Katz_working_2021}
D.~S. Katz, F.~Psomopoulos, and L.~J. Castro, ``\BIBforeignlanguage{en}{Working {Towards} {Understanding} the {Role} of {FAIR} for {Machine} {Learning}},'' in \emph{\BIBforeignlanguage{en}{Proc. 2nd Workshop on Data and Research Objects Management for Linked Open Science}}, Virtual Conference, 2021, pp. 1--6.

\bibitem{schwardmann_automated_2016}
U.~Schwardmann, ``Automated schema extraction for {PID} information types,'' in \emph{Proc. {IEEE} {International} {Conference} on {Big} {Data} ({Big} {Data})}, Dec. 2016, pp. 3036--3044.

\bibitem{kahn_framework_2006}
R.~Kahn and R.~Wilensky, ``\BIBforeignlanguage{en}{A framework for distributed digital object services},'' \emph{\BIBforeignlanguage{en}{Int. J. Digit. Libr.}}, vol.~6, no.~2, pp. 115--123, 2006.

\bibitem{doip_spec}
\BIBentryALTinterwordspacing
{DONA Foundation}, ``{Digital Object Interface Protocol Specification},'' 2018. [Online]. Available: \url{https://www.dona.net/sites/default/files/2018-11/DOIPv2Spec_1.pdf}
\BIBentrySTDinterwordspacing

\bibitem{cousijn_connected_2021}
H.~Cousijn, R.~Braukmann, M.~Fenner, C.~Ferguson, R.~van Horik, R.~Lammey \emph{et~al.}, ``\BIBforeignlanguage{eng}{Connected {Research}: {The} {Potential} of the {PID} {Graph}},'' \emph{\BIBforeignlanguage{eng}{Patterns}}, vol.~2, no.~1, p. 100180, Jan. 2021.

\bibitem{broeder_data_2014}
D.~Broeder and L.~Lannom, ``Data {Type} {Registries}: {A} {Research} {Data} {Alliance} {Working} {Group},'' \emph{D-Lib Mag.}, vol.~20, 2014.

\bibitem{schwardmann_digital_2020}
U.~Schwardmann, ``Digital {Objects} – {FAIR} {Digital} {Objects}: {Which} {Services} {Are} {Required}?'' \emph{Data Sci. J.}, vol.~19, p.~15, 2020.

\bibitem{schultes_fair_2019}
E.~Schultes and P.~Wittenburg, \emph{{FAIR} {Principles} and {Digital} {Objects}: {Accelerating} {Convergence} on a {Data} {Infrastructure}}.\hskip 1em plus 0.5em minus 0.4em\relax Cham: Springer International Publishing, 2019, pp. 3--16.

\bibitem{doi/10.2777/620649}
E.~Commission, D.-G. for Research, Innovation, O.~Corcho, M.~Eriksson, K.~Kurowski \emph{et~al.}, \emph{EOSC interoperability framework – Report from the EOSC Executive Board Working Groups FAIR and Architecture}.\hskip 1em plus 0.5em minus 0.4em\relax Publications Office, 2021.

\bibitem{deeleman-reinhold_genera_2024}
C.~Deeleman-Reinhold, W.~Addink, and J.~Miller, ``\BIBforeignlanguage{en}{The genera {Chrysilla} and {Phintelloides} revisited with the description of a new species ({Araneae}, {Salticidae}) using digital specimen {DOIs} and nanopublications},'' \emph{\BIBforeignlanguage{en}{Biodivers. Data J.}}, vol.~12, p. e129438, Sep. 2024.

\bibitem{islam_assessing_2023}
S.~Islam, J.~Beach, E.~Ellwood, J.~Fortes, L.~Lannom, G.~Nelson \emph{et~al.}, ``Assessing the {FAIR} {Digital} {Object} {Framework} for {Global} {Biodiversity} {Research},'' \emph{Res. Ideas Outcomes}, vol.~9, Sep. 2023.

\bibitem{mayer_thermal_2023}
Z.~Mayer, J.~Kahn, M.~Götz, Y.~Hou, T.~Beiersdörfer, N.~Blumenröhr \emph{et~al.}, ``Thermal {Bridges} on {Building} {Rooftops},'' \emph{Sci. Data}, vol.~10, no.~1, p. 268, 2023.

\bibitem{bonino_da_silva_santos_towards_2023}
L.~O. Bonino~da Silva~Santos, T.~P. Sales, C.~M. Fonseca, and G.~Guizzardi, ``Towards a {Conceptual} {Model} for the {FAIR} {Digital} {Object} {Framework},'' in \emph{Formal {Ontology} in {Information} {Systems}}.\hskip 1em plus 0.5em minus 0.4em\relax IOS Press, 2023, pp. 227--241.

\bibitem{soiland-reyes_evaluating_2024}
S.~Soiland-Reyes, C.~Goble, and P.~Groth, ``\BIBforeignlanguage{en}{Evaluating {FAIR} {Digital} {Object} and {Linked} {Data} as distributed object systems},'' \emph{\BIBforeignlanguage{en}{PeerJ Comput. Sci.}}, vol.~10, p. e1781, Apr. 2024.

\bibitem{mayer_deep_2023}
Z.~Mayer, J.~Kahn, Y.~Hou, M.~Götz, R.~Volk, and F.~Schultmann, ``Deep learning approaches to building rooftop thermal bridge detection from aerial images,'' \emph{Autom. Constr.}, vol. 146, p. 104690, 2023.

\bibitem{krausGoldStandardBenchmark2024}
F.~Kraus, N.~Blumenr{\"o}hr, G.~G{\"o}tzelmann, D.~Tonne, and A.~Streit, ``A {{Gold Standard Benchmark Dataset}} for {{Digital Humanities}},'' in \emph{Proc. of the 19th {{International Workshop}} on {{Ontology Matching}}}, Baltimore, USA, in press.

\bibitem{jejkal_realizing_2022}
T.~Jejkal, A.~Pfeil, J.~Schweikert, A.~Pirogov, P.~Barranco, F.~Krebs \emph{et~al.}, ``Realizing {FAIR} {Digital} {Objects} for the {German} {Helmholtz} {Association} of {Research} {Centres},'' \emph{Res. Ideas Outcomes}, vol.~8, p. e94758, Oct. 2022.

\bibitem{Stocker-2020}
M.~Stocker, L.~Darroch, R.~Krahl, T.~Habermann, A.~Devaraju, U.~Schwardmann \emph{et~al.}, ``Persistent identification of instruments,'' \emph{Data Sci. J.}, vol.~19, p.~18, May 2020.

\bibitem{schwardmann_how_2023}
U.~Schwardmann and T.~Kálmán, ``\BIBforeignlanguage{en}{How {FDO} attributes can support machine- and human-readability? - a description along three examples},'' \emph{\BIBforeignlanguage{en}{Res. Ideas Outcomes}}, vol.~9, p. e108737, Oct. 2023.

\end{thebibliography}

\end{document}